# A Mathematical Framework for Reading the *Autopsias'* Meta-Compositional System

Patricio F. Calatayud [1], Pablo Padilla Longoria [2], Álvaro Martínez Ramírez [3]

[1] *Phd music program, UNAM.* patricio.tics@gmail.com

[2] *Phd music program, UNAM.* pabpad@gmail.com

[3] *Undergraduate, faculty of Science, UNAM.* alvaro.mtr@gmail.com

## Abstract

Background: New forms of music writing using computers have arisen in the past 20 years. Most of them use the capacities of digital manipulation of data like animation, algorithmic processing, cinematic view, and much more. These scores use dynamic musicography, and all of them share a problem: They have readability problems. We will argue that this problem can be addressed by mathematical tools.

Aims: Take the *Autopsias'* [Autopsies] meta-compositional system as a study case for starting the construction of a mathematical framework that can overview the readability for dynamic musicography. The *Autopsias'* system is the process of transforming a musical score dynamically.

Methodology: We will start to build a mathematical framework by taking a group of basic topological concepts, and applying them after a bridge between a printed score and a dynamic computational re-appropriation of it has taken place. We will observe the writing and the performing of the several performances of the *Autopsias'* composition, and analyze them from a mathematical perspective, in order to get the ability to start constructing a new set of orthographical tools for the performance of dynamic musicography.

Main Contribution: The performances of the *Autopsias'* meta-compositional system show that mathematics, as with pitch-class sets, can be helpful for understanding music information, as well as its performance. Also show the necessity of augmenting the set of orthographic rules, in order to improve the dialog between composers and performers.

Conclusion and Implications: This initial approach concludes that there exists a relation between the original interpretation of a score, and the mathematical re appropriation of it. The mathematical systematization of the qualia of music, beyond the music pitch, could be useful for generating a renewed readability abilities for the composer that wants to use other forms of music notation.

*A sign is said to be iconic when there is a topological similarity between the signifier [the sign] and its denotata [what it represents]* (In Roads 1985: 405; Sebeok 1975: 242)

# 1. Introduction

Digital music is being part of a major transformation that involves programming music scores and its screen projection. After more than 60 years of computer music a number of composers changed their methods in order to re-obtain sounds from human musicians –instead of electronic sounds. They started building dynamical notation and the scores generated by this manipulation have been named in different ways, however the term 'dynamic musicography' designates its two main characteristics: They exhibit dynamical behavior and change their contextual operability –their performance. They are animated, automatically generated, they also change according to the environment, modified by internal or external feedback, etc. Also these scores will no longer be the private space of a musician or an orchestral conductor; they will be projected to a potentially evaluating audience. Then the question arises: Can we read non-static notation?

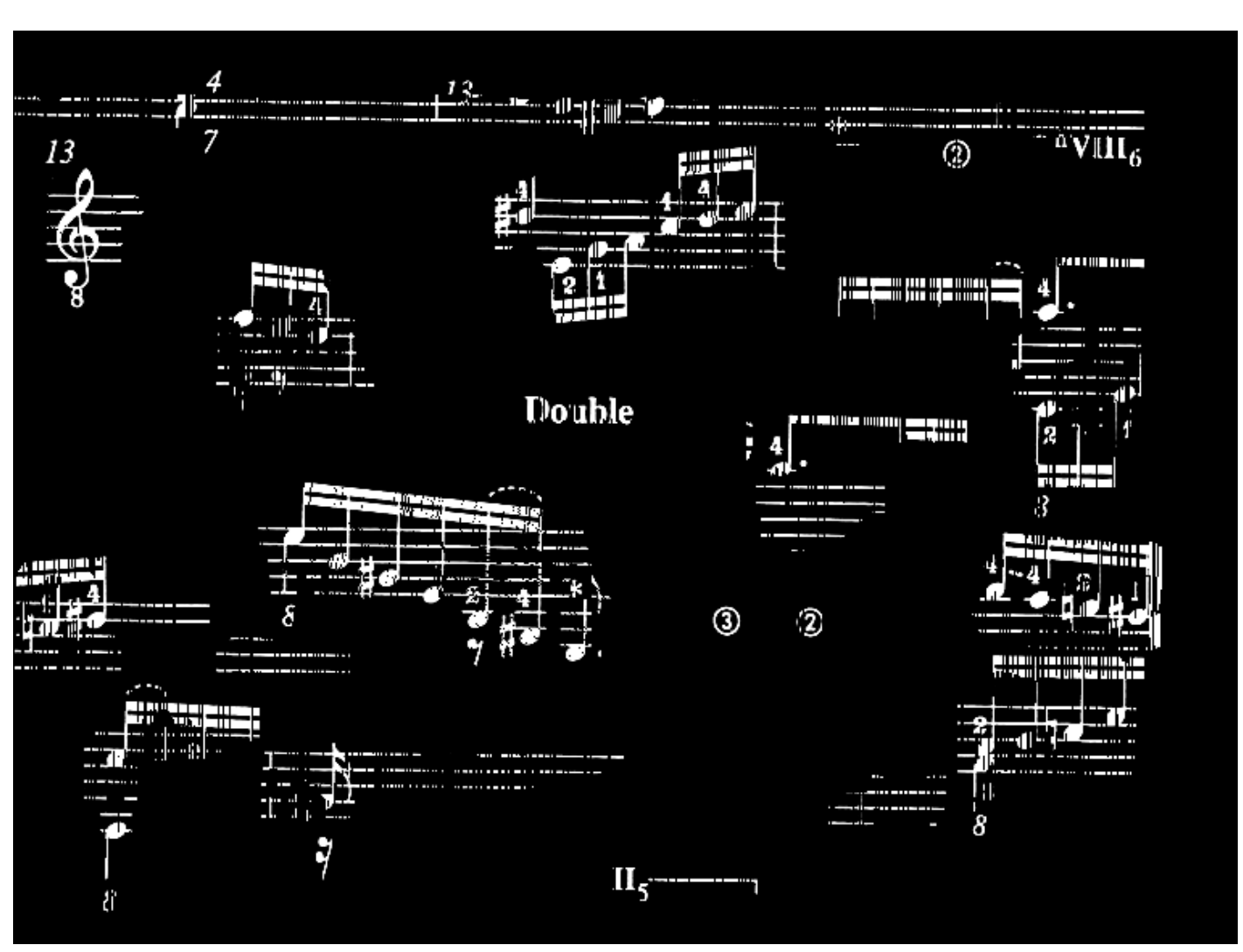

Figure 1: *Autopsia* applied to *Partita*, op. 997, '*Double*' from J.S. Bach (1980). An only visual version of it can be found at https://youtu.be/I_e6rBZMuaQ, and an instrumental one at https://youtu.be/xJGnyTgTMGE (10-03-2020)

Musicians who play dynamic musicography do not know what to do when a notation is displaced from its fixed or 'natural' position. They also claim a readability problem that we will prove that can be addressed with formal mathematical tools. Music and mathematics go a long way, if we use this discipline to organize notation rather than sound, we will be able to understand readability in music notation. Building a framework with those tools will lead to a better notion of design, towards the generation of a readable medium. A similar use of this mathematical approach has been taken by Aarseth (1997): 1. Distinguishing a score (as a constructed object), 2. Its writing and reading (as temporal processes), and 3. understanding the lines and nodes of hypertexts accessibility as a matter of topology rather than tropology (Aarseth 1997, 42–44).

Now, if we can generalize any rule with mathematical tools, we could ask: Can we measure music reading using mathematics? We will be able to answer these questions using a musical piece that: a) Is possible to be read with grammatical and orthographic rules; b) Has elements that can be observed from mathematical designs; and c) These elements are in motion. For this reason we will use the *Autopsias'* meta-compositional system (a composition that builds compositions) as a study case.[1] We will divide the processes of the *Autopsias'* performance in writing, and reading. In this way we will be able to prove that, despite a geometric inconsistency in the transformation of a score, the performance of it shows similarities. This outcome is a first step in analyzing the performance of dynamic musicography.

Our approach, even though it's built by mathematical tools, will be similar to those basic and intuitive approaches like the ones of Lacan (Amster 2011), and of some musicians that start the connection with these complex theories. Topological tools have already been applied to music (Mazzola 2012; Tymoczko 2010), we need to expand them beyond the 'Pitch Class Set' theory to include a spatial analysis of dynamic musicography. This action will help to develop a sort of orthography, according to the new computer capacities to generate scores.

While an *Autopsias'* is performed, the elements of a original score is transformed. We want to know whether mathematical concepts, like topology and graph theory, can be used as tools to describe transformations in musical notation and its readability.

---

[1] We will speak about the piece later on.

Conceptually, topology studies the collection of progressive transformations in shapes: surfaces, knots, and their qualitative consequences. Those objects that remain the same are called ‘invariants’, while the others will be considered as ‘tears’ from the original image. It will be useful to compare the musical performance between one printed score and a dissected version of it –built with the *Autopsias*' system. Topology is also aware of the cracks, tearings and voids in those structures, and in this text we will distinguish tears in the original score and tearings in the reading of it.

## 2. The *Autopsias'* system

An *Autopsia* is a process of intervention of the gestures involved in the performance of a score. The author of the composition takes a score, provided by a performer, and transforms it, in real time, projected on a screen, and in front of an audience. In order to play any version of the *Autopsias*’ system we need a musician (or an established group of musicians) to provide a printed (or digital) music score from their repertoire. We also need a computer musician that builds the score in real time with a computer and projects it. After a dialog between the musicians involved, and a creative and collaborative process is done, a list of fragments is created –a sort of a renewed lexicon of measures, text, or even printed noise.

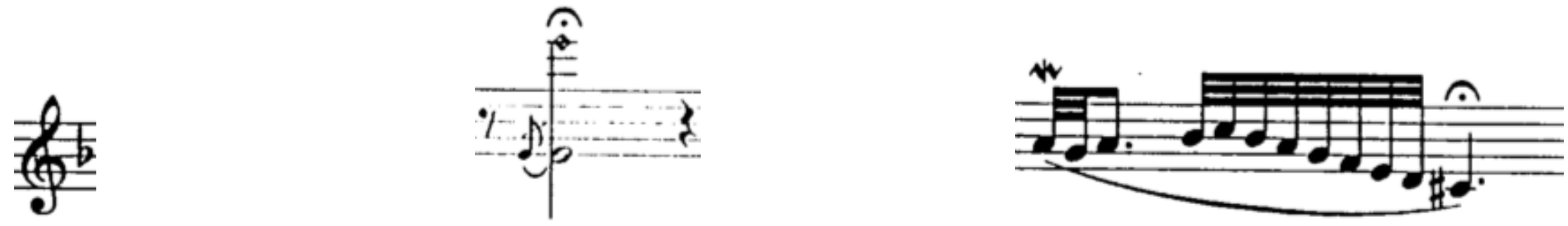

Figure 2, 3, and 4: Three selected fragments from 3 *virtuoso Flamenco Studies* by Krystof Zgraja (1996).

These will be re-appropriated and reorganized, spatially and semiotically, in a pictorial way. This performance is made during rehearsals and live on stage, always giving a different version of the score, and its performance.

Having a realm R as our original score and a realm A as the Autopsias’ score, the musicians will perform this piece with (mainly) two possible outcomes:

- An ‘identical’ (intrinsic, **R = A**) one, made possible by sight reading, where they will play the fragments in the same way as they played them before –remembering how the original score was played in a previous rehearsal. From this literal (prosodic) reading, and the feedback from his or the other player’s sounds, music will emerge.

- An ‘equivalent’ (extrinsic; **R ≈ A**) one, where the sounds produced, instead of reflecting the displayed score, are improvisational variations of the original. As this type of outcome takes place, an improvisational system is being built; initially activated thanks to a reaction to the content displayed on the screen. This outcome can be understood in a different scheme that includes all the re-organized fragments of the original score –expanding the edges of the first outcome, as we will see later. This performance of the *Autopsias*' score will result in an improvisational or comprovisational process, conditioned by the musician’s background of their previous performance.

Why is this study topological and not only geometrical? Because a teared score does not imply a teared performance of it. If we improvise in a tonal way, and we only play with four of the seven notes of the scale, we are tearing apart the tonal system?

In this video we can see the process of building a version of *Autopsia*, made for the appropriation of *Tristes hombres si no mueren de amores* by Leo Brouwer (1987).

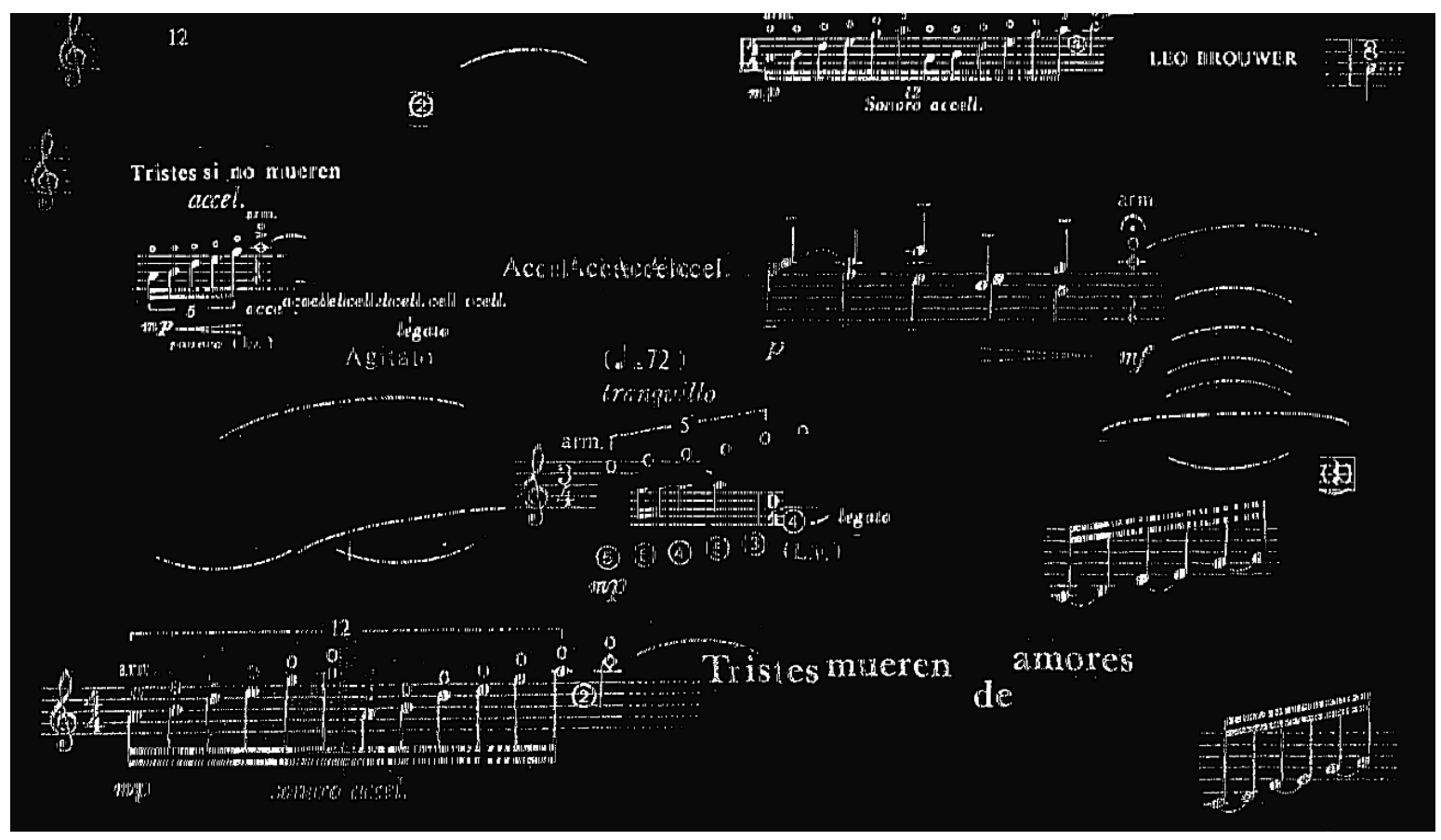

Figure 5: *Autopsia* applied to the *Tristes hombres si no mueren de amores...,* by Leo Brouwer (1987). An instrumental version can be found at https://youtu.be/Cnw063Tb3nQ (10-03-2020)

If we are going to perform the outcomes of the *Autopsias*' system, we can use cognition tools and a mathematical analysis for understanding its writing and performance. Together, these two disciplines can build a framework that includes improvisational and comprovisational decisions, in order to fully understand the new potential in the performance of dynamic musicography.

## 3. Performance scheme in *Autopsias'*

The performance of the *Autopsias'* score is neither conducted by pitches (like tonal improvisation), music style (as flamenco), nor absolute freedom (like Indeterminacy), it is conditioned by his previous performance of the original score. Similarly to the previous question, if we improvise with three notes of a raga, are we tearing it apart? Or on the contrary, how many ornaments blur the (mathematical) identity of it?

In general, the scores that result from *Autopsias'* system can be seen as pictorial lead sheets built in real time. To exemplify this, lets see the following:

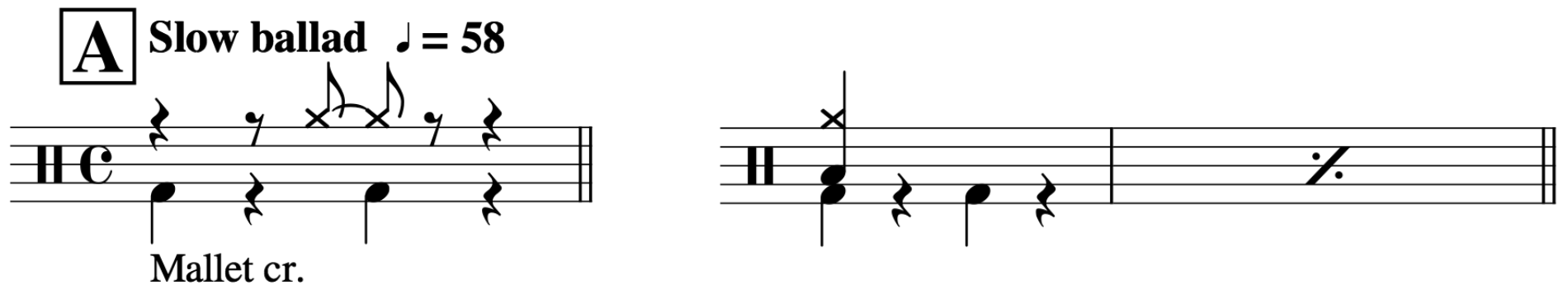

Figures 6 and 7: Two fragments from *Blood Count* by Billy Strayhorn ®1979 ASCAP.

In jazz performance we never play two measures the same way. The expansion that generates those motives can be described like the infinite quality of a geometric line. Even if a line generates curves or knots, it is still the same line –topologically speaking. Does the riffs from the example sound the same in another song?[2] We could even ask if it sounds the same in several 'takes' of the same song. In this sense we cannot build a topological distinction by accumulating, and collecting musical objects; instead, topology helps separating an intrinsic view –the musical fragment itself–, and an extrinsic view –where and when this fragment is being played.

[2] We take a 'riff' as a musical pop phrase or gesture that is repeated *ad* lib.

## 4.a Writing the score in *Autopsias'*

The *Autopsias* system is performed in real-time. It starts from an empty screen that is progressively filled with notation from another score. As it is built, it is performed by a group of musicians.

Previously, after a dialog with the performers, a set of 'morphemes' (Stroppa 1999) is chosen, and they can be named 'symbol' with an underscore to index them.

$$S_1 = \langle t, p, d \rangle \quad (1)$$

where t, p, d, short for time, pitch, and duration, are the parameters that we will transform in the new score. The complete sucesion of these symbols, we denote it by

$$V = \langle s_1, s_2, \ldots, s_n \rangle \quad (2)$$

Now, when this collection is placed in the new visual structure, we must understand them as a set, where the elements are organized in a new way (randomlike),

$$A = \{\mathcal{J}(s_1), \mathcal{J}(s_2), \ldots \mathcal{J}(s_n)\} \quad (3)$$

understanding that the A is a subset V,

$$A \subset V \quad (4)$$

because all elements from V are seen in A,

$$s_n \in A \Rightarrow s_n \in V \quad (5)$$

With the writing process of *Autopsias*' score we will get a morpheme with a new timbral quality. The repercussion of these changes (eg. the intuitive set of rules that determine what to play if a morpheme changes) are established beforehand between the builder of the score, and the performer.

## 4.b Performing the score in *Autopsias'*

This process can be understood by the mathematical notion of finite automata, which is essentially a directed graph describing the evolution of a 'state'. In this concept, a directed graph is a pair of two sets; one is the collection of 'vertices' or states, and the other the collection of arrows between these states.

$$G = \langle \Omega, E \rangle \quad (6)$$

The fragments of dissected original are the vertices,

$$\Omega = \{chosen\ fragments\ of\ the\ original\ score\} \quad (7)$$

reorganized in space and linked by proximity of meaning (arrows).

$$E = \{\langle u, v \rangle \mid u, v \in \Omega\ and\ v\ can\ be\ performed\ after\ u\} \quad (8)$$

That is: the set of arrows E, is a collection of pairs of vertices. Each pair has the initial and the final which means that an arrow $a \in E\ a = \langle u, v \rangle$begins in u and ends in v.

When we read the original score, and perform it to build a music scheme we do it from left to right, from top to bottom. We can describe this scheme with the mathematical notion of finite automata (Hopcroft & Ullman 1979).

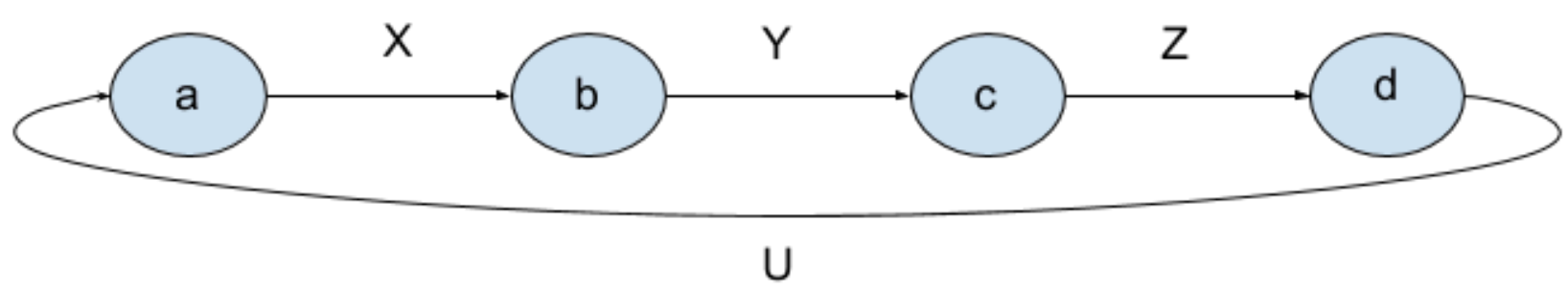

Figure 9: A finite automata scheme for the reading of the original score.

With the same notation as before, the vertices are {a,b ,c, d} and the arrows {X, Y, Z,U}

$$G_{orig} = \langle \{a, b, c, d\}, \{X, Y, Z, U\} \rangle \quad | \quad (abcd)* \quad (9)$$

and the only phrase that we can build is $abcd$ and it can be repeated indefinitely (hence marked with the ‘*’).

But the morphemes that are relocated in the performance of *Autopsias* do not necessarily follow the same scheme. We could find something like this,

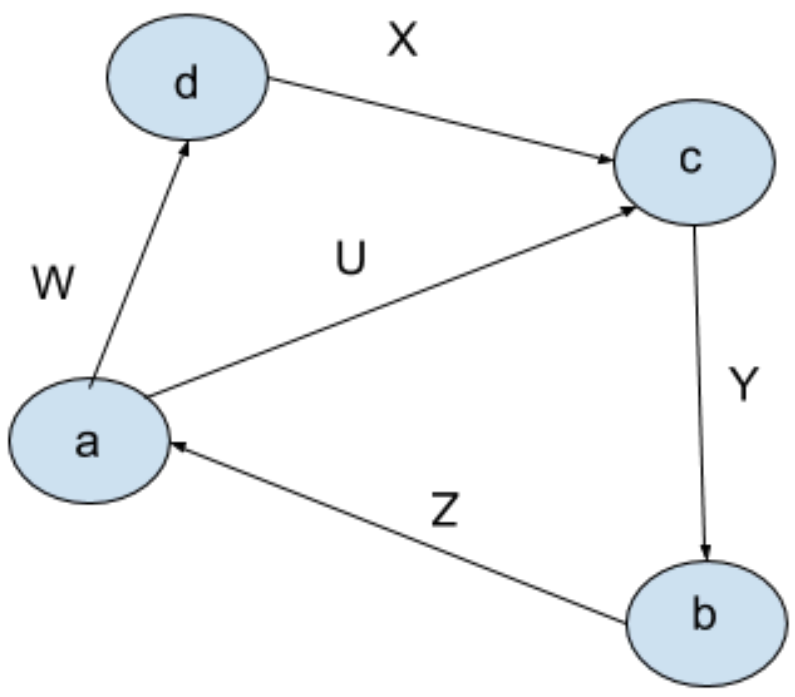

Figure 10: An alternative finite automaton scheme that describes the reading scheme for the same morphemes that before, but manipulated by the *Autopsias*’ system.

Now here the word possibilities increase, and it can be notated in this way,

$$G_{aut} = \langle\{a, b, c, d\}, \{X, Y, Z, W, U\}\rangle \mid (a(d)?\, cb)\ * \qquad (10)$$

This scheme could also allow a reading with this denotation,

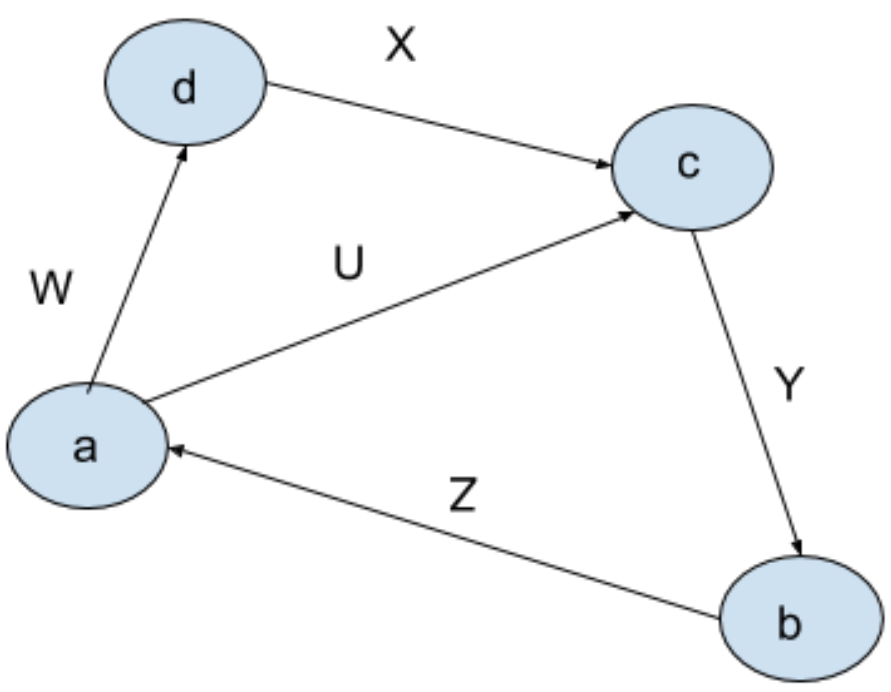

Figure 11: An alternative scheme for the same finite automata.

On the other hand, when we transform the morpheme location, we can also change another parameter (see (1)). The way that we can do this is by applying a function on each one of its elements.

$$F_{autopsia}(S_i) = \langle F(s_1), F(s_2), F(s_3) \ldots F(s_n) \rangle, \qquad (11)$$

where the $F$ function can change each of the elements of the morpheme.

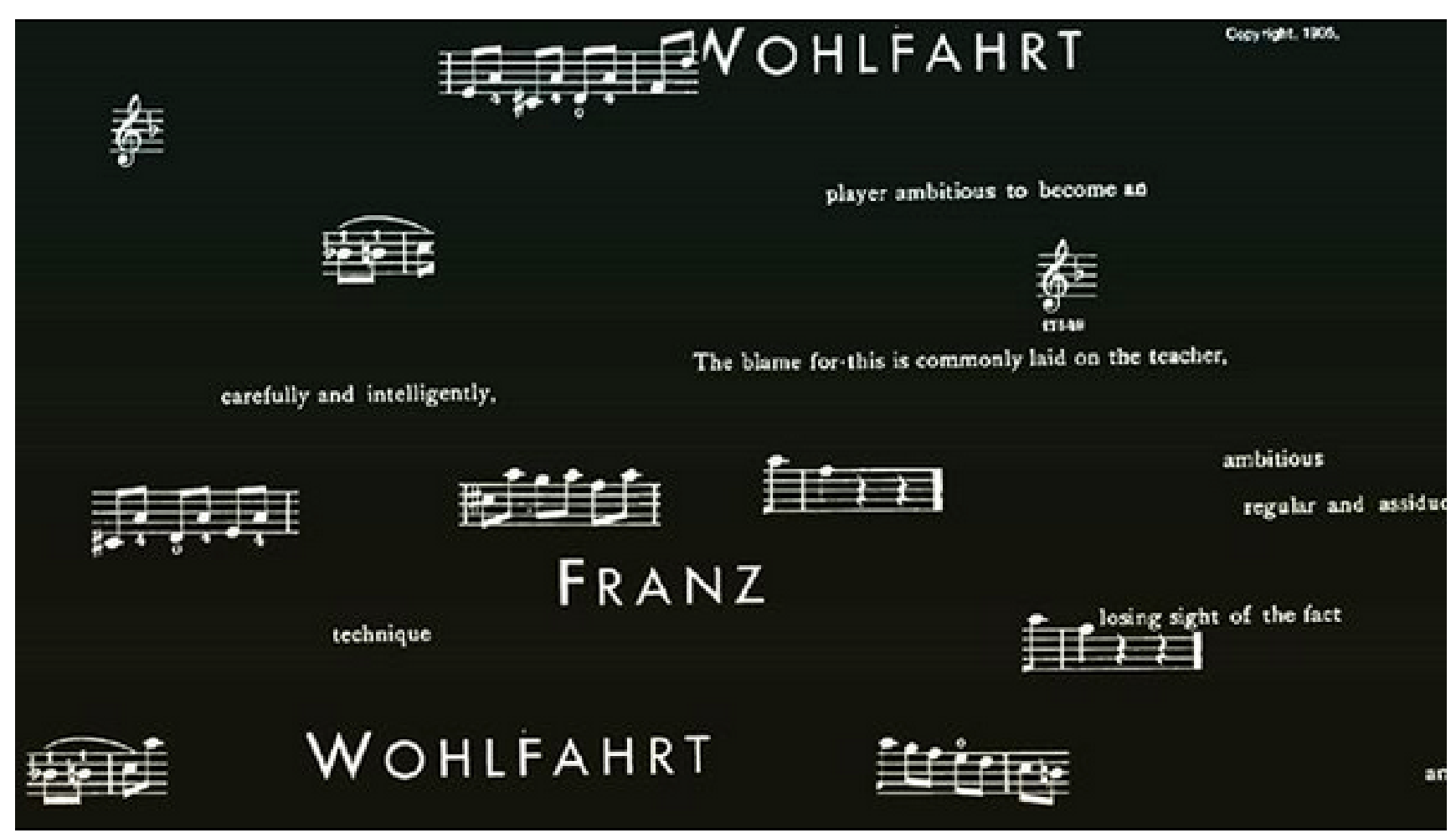

Figure 12: *Autopsia* applied to the Study for violin no.23, 'Moderato' by Franz Wolfhart (1905).

# 5. Conclusions and future work

The relationship between the original score and the sounds coming out of its reconstruction depends on complex concepts that can be understood by mathematics. In this sense, mathematics and music continue to relate, now in a new direction that tries to organize new notation instead of new sounds. The readability scheme of dynamic musicography that results from this approach also renews the orthographic context of music texts, allowing new studies to take advantage of other disciplines like music cognition tools.

Under the scope of topology, we can observe changes between the $G_{aut}$ set and the $G_{origt}$ one. The point of this work is to offer the musicians –at least the ones that perform this work–, an analysis framework for a better readability and performance of the system.

Another approach could be the integration of 'Temporal Gestural Units', developed by Tenney and Polansky (Tenney and Polansky 1968). This approach goes beyond the mathematical analysis of pitch, and its tonal relations in order to make an analysis of space and meaning.[3] Also, splitting fragments of a score, more than continuing with its etymology can be understood as a decontextualization, that changes the dynamics of the original reading. This is similar to the reorganization of words of a given text with the objective to build sonic poetry.

We will also need to study the performer's intention to understand the degree of freedom that they somehow overwrite over the composer's one. This is clear in the several performances of the *Autopsias's* meta-compositional system.

Lastly, let's remember that actual orthography is linked to a printing device. (Quignard 2016) Dynamic musicography suggests the development of a new form of orthography that involves dynamic notation and spacing placement into account.

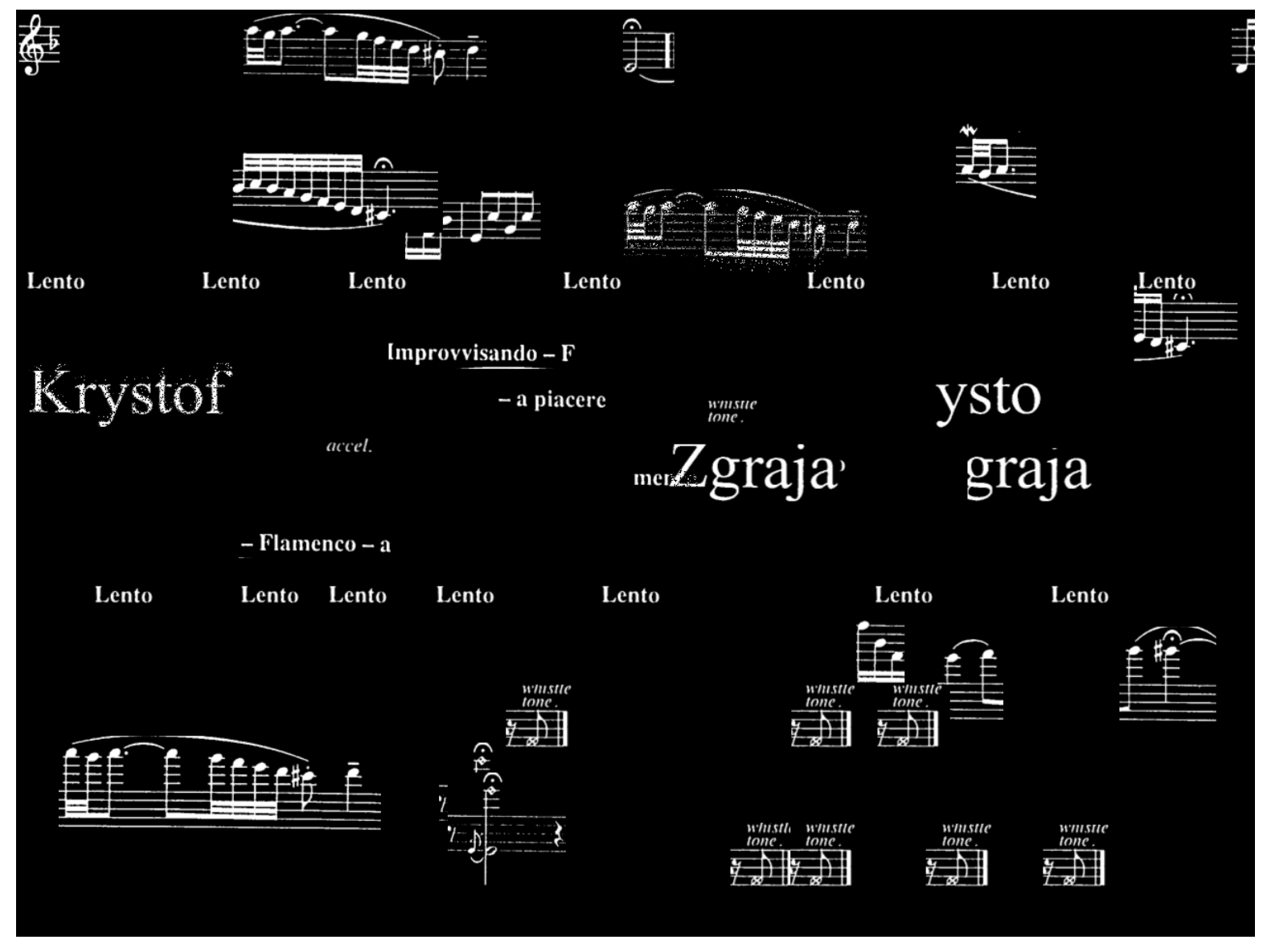

Figure 13: *Autopsia* applied to 3 virtuoso Flamenco Studies by Krystof Zgraja (1996). A live version is at https://youtu.be/GB0inAL90wg (10-03-2020)

[3] We will need to change 'morphemes' into 'clangs'. But this analogy also takes the name of 'object sonore', 'macrotimbre', 'musical types', and many others.

## Acknowledgments

Pablo Amster & Bruno Mesz for their mathematical patience, and María Teresa Campos Arcaraz for the help with edition and recommendations.